\PassOptionsToPackage{sort&compress}{natbib}
\documentclass[onecolumn,numbers]{els-mrw} 

\usepackage{amsmath,amssymb,amsfonts,amsthm,makeidx,graphicx}
\usepackage{txfonts}
\usepackage{helvet}
\usepackage{braket}
\usepackage[colorlinks=true,citecolor=blue,linkcolor=blue,urlcolor=blue]{hyperref}

\usepackage{orcidlink}

\newcommand{\fmi}{\ensuremath{\mathrm{fm}^{-1}}}
\newcommand{\mev}{\ensuremath{\mathrm{MeV}}}

\begin{document}

\chapter{Similarity renormalization group for nuclear forces}\label{chap1}

\author[1,2]{Matthias Heinz\orcidlink{0000-0002-6363-0056}}

\address[1]{\orgname{Oak Ridge National Laboratory}, \orgdiv{National Center for Computational Sciences}, \orgaddress{Oak Ridge, TN 37831, USA}}
\address[2]{\orgname{Oak Ridge National Laboratory}, \orgdiv{Physics Division}, \orgaddress{Oak Ridge, TN 37831, USA}}

\maketitle

\begin{abstract}
    Renormalization group methods generate low-resolution Hamiltonians that are more diagonal, with reduced coupling between low- and high-energy states, and thus easier to solve.
    This chapter reviews the similarity renormalization group for nuclear Hamiltonians,
    which is a popular method for generating low-resolution nuclear forces.
    It presents the similarity renormalization group flow equations, analyzes how the similarity renormalization group drives the Hamiltonian towards the diagonal, and studies the effect of induced many-body interactions.
    It concludes by highlighting the progress in first-principles calculations of nuclei driven by low-resolution nuclear Hamiltonians.
\end{abstract}

\begin{keywords}
  	Nuclear forces \sep renormalization group methods \sep similarity renormalization group \sep low-resolution potentials \sep first-principles nuclear theory
\end{keywords}

\begin{glossary}[Nomenclature]
\begin{tabular}{@{}lp{34pc}@{}}
EFT & effective field theory\\
QCD & quantum chromodynamics \\
RG  & renormalization group\\
SRG & similarity renormalization group\\
\end{tabular}
\end{glossary}

\section*{Key points}
\begin{itemize}
\item Hamiltonians often have have significant off-diagonal matrix elements arising from short-distance interactions when represented in momentum space or an energy-ordered basis, making them nonperturbative and difficult to solve.
\item Renormalization group methods transform Hamiltonians to lower resolution, replacing explicit, complicated short-distance physics with effective low-energy operators. Such low-resolution potentials are more perturbative and easier to solve.
\item The similarity renormalization group generates a continuous unitary transformation, which may be tailored to drive the Hamiltonian towards a diagonal form.
\item The similarity renormalization group applied to nuclear forces systematically generates low-resolution potentials that are unitarily equivalent to the original Hamiltonians.
The transformation also induces three-body and higher-body forces, which must be treated to achieve exact renormalization group invariance in observables.
\item Low-resolution nuclear potentials have driven progress in first-principles nuclear theory due to their perturbativeness and fast convergence behavior.
  
\end{itemize}

\section{Introduction}

Modern nuclear theory has been defined by our developing understanding of theoretical resolution and how to exploit it in studies of nuclear forces, atomic nuclei, and nuclear matter.
At its core, theoretical resolution is a choice we make about what to do with high-energy, short-wavelength physics.
This short-wavelength or short-range physics may be complicated, difficult to compute, and potentially completely unnecessary to treat explicitly in describing a phenomenon of interest, for example, a nuclear ground state or low-energy scattering.
Working at the appropriate resolution means only essential low-energy, long-wavelength physics is included explicitly while the remaining unresolved short-range physics is captured effectively.

While this concept may seem intuitively simple, its application is not always straightforward.
Shifting to lower theoretical resolution often requires switching to effective low-energy degrees of freedom and determining unknown effective low-energy interactions.
Effective field theories~(EFTs)~\citep{Weinberg1979PASMA_EFT, Epelbaum2009RMP_ChiralEFTReview, Machleidt2011PR_ChiralEFTReview, Hammer2020RMP_NuclearEFT} and renormalization group~(RG) techniques~\citep{Wilson1974PR_RenormalizationGroup, Bogner2010PPNP_RGReview} provide systematic ways to construct Hamiltonians at a specific resolution.
Both methods have been broadly successful in many domains of physics, 
and together they have spurred tremendous progress in nuclear theory in the past 25 years.
EFTs allow us to construct a low-resolution theory with effective degrees of freedom that is systematically rooted in a higher-resolution underlying theory.
Renormalization group methods, on the other hand, allow us to adjust the resolution by integrating out or decoupling irrelevant short-range physics.
These methods work well together: 
We can use an EFT to build a Hamiltonian at a specific resolution and then use RG methods to further tune the resolution of the Hamiltonian, if necessary.

This chapter focuses on the similarity renormalization group (SRG)~\citep{Glazek1994PRD_SRG, Wegner1994AP_SRG, Bogner2007PRC_SRGNN}.
The SRG is a remarkably simple approach:
It is an operator differential equation to compute a unitary transformation of the Hamiltonian, typically used to evolve the Hamiltonian into a more diagonal form.
In its simplicity, it provides a lot of flexibility, and the SRG has been used to develop low-resolution Hamiltonians and also to solve the quantum many-body problem in fields ranging from cold atoms to quantum chemistry to nuclear physics.
Here, we focus on using the SRG to construct low-resolution nuclear forces.
We start by revisiting the concept of resolution in detail and making the benefit of low-resolution Hamiltonians clear.
The next section introduces the similarity renormalization group, exploring how the RG transformation works and how it impacts the structure of the Hamiltonian.
After this, we focus on the application of the SRG to nuclear forces,
discussing some key lessons learned over the past 20 years and some challenges that are not yet fully addressed.
The last section highlights the progress in computations of nuclei made possible by the SRG and low-resolution Hamiltonians.

\section{Physics at low resolution and the renormalization group}

% Let us start by revisiting the concept of resolution.
In the modern world, resolution is often understood as digital resolution, the density of pixels in an image, which determines the minimum scale of details that image can capture.
In physics, one often also thinks about experimental resolution, which is determined not only by the resolution of the instruments one is using, but also by the energy at which the system is probed.
High-energy probes have a shorter de Broglie wavelength, which in turn allows them to resolve finer details.
Theoretical resolution is analogous: A state $\ket{n}$ with energy $E_n$ resolves interactions with characteristic momentum scales $Q$ with kinetic energies $Q^2/(2\mu)\lesssim E_n$ (with the reduced mass of the system $\mu$).
Getting the microscopic description of such interactions right is important to correctly model the state $\ket{n}$.

Conversely, it first seems plausible that modeling a low-energy state, for example, the ground state $\ket{\psi}$ with energy $E_\mathrm{gs}$, does not resolve interactions with $Q^2/(2\mu) > E_\mathrm{gs}$.
However, it is easy to see that this is not necessarily the case:
If one considers the expansion of the ground-state energy in perturbation theory
with the unperturbed Hamiltonian $H_0$ with eigenstates $\ket{n}$ and the perturbation $H_1$,
the second-order correction is
\begin{equation}
    \label{eq:pt2}
    E^{(2)} = \sum_{n\neq 0} \frac{|\braket{0 | H_1 | n}|^2}{E_0 - E_n}\,.
\end{equation}
Clearly, there is an unrestricted sum over virtual states $\ket{n}$, including many with $E_n \gg E_0 \sim E_\mathrm{gs}$.
Whether these high-energy states give significant contributions to the ground-state energy depends on the magnitudes of the matrix elements $\braket{0| H_1 |n}$ relative to those of the energy differences $E_0 - E_n$.
In other words, if $H_1$ has strong couplings between $\ket{0}$ and $\ket{n}$ for $E_n \gg E_0$, those virtual states are important.
Accordingly, the description of the ground state actually requires higher resolution than one might expect.
On the other hand, if high-energy states $\ket{n}$ are decoupled from $\ket{0}$, meaning the matrix elements $\braket{0| H_1 |n}$ are small,
the sum in Eq.~\eqref{eq:pt2} may be truncated at small $n$.
This means the ground state can be expanded in low-energy states $\ket{n}$ with $E_n$ less than some energy scale $\Lambda$,
which is the intrinsic resolution scale of the Hamiltonian $H_1$.

Thinking from the perspective of theoretically predicting low-energy phenomena, we can clearly see two very distinct scenarios.
When the Hamiltonian resolution scale $\Lambda$ is large compared to the typical momentum scale $Q$,
our states will be very complicated, with significant contributions from high-energy virtual states.
The second-order correction in Eq.~\eqref{eq:pt2} will need to be evaluated in a large basis.
Moreover, higher-order corrections in perturbation theory will also be significant, and, in general, perturbation theory will diverge and fail,
meaning we need to employ nonperturbative approaches.
If instead $\Lambda$ is comparable to $Q$,
things are much simpler. 
A small basis of a few $\ket{n}$ is sufficient to converge Eq.~\eqref{eq:pt2},
and perturbation theory is better behaved.
Clearly, if possible, we would prefer a lower resolution scale.

So what actually determines the resolution scale $\Lambda$? The answer is: We choose the resolution scale, either implicitly or explicitly.
First, we choose the degrees of freedom we use to describe a system.
For strong-interaction systems, for example, the fundamental degrees of freedom are quarks and gluons, and their interactions are given by quantum chromodynamics (QCD).
In nuclei, however, all quarks are confined in colorless protons and neutrons, so nucleons are more appropriate degrees of freedom.
In this case, the complicated dynamics of QCD are replaced by effective nuclear forces between nucleons.
What are these nuclear forces?
Again, we make a choice here.
We can use phenomenological models, as was popular in the previous century, or instead we can use effective field theories and renormalization group techniques to systematically construct nuclear Hamiltonians at a specific resolution.

The freedom to choose the appropriate degrees of freedom and the appropriate (effective) interactions,
recasting intractable high-resolution theories into tractable low-energy effective theories,
has been essential for progress in theoretical physics.
The renormalization group in particular has played a key role in understanding and systematizing the development of effective theories.
Many different variants of RG methods exist, but all share some common features:
\begin{enumerate}
    \item The RG transformation systematically changes the resolution scale. In low-energy nuclear physics, it typically reduces the resolution scale, integrating out high-energy states and degrees of freedom or decoupling them from those relevant at the low-energy scale of interest.
    Transforming to lower resolution replaces explicit short-range interactions with effective low-energy interactions
    and short-range degrees of freedom with collective degrees of freedom.
    \item Low-energy observables must be preserved, meaning that the short-distance physics that has been decoupled must be replaced with effective low-energy operators.
    These are resolution dependent (often also referred to as scale dependent) and change continuously as the RG transformation is performed.
    Renormalization group invariance (or at least approximate invariance) of low-energy observables is a key diagnostic of the method.
    \item RG transformations to low resolution also induce many-body interactions.
    Many-body interactions are natural in effective theories with composite low-energy degrees of freedom
    (for example, three-body forces when modeling the sun-Earth-moon system using point-like masses).
    RG transformations show that the division between two-, three-, and higher-body forces is resolution dependent.
    \item When transforming to very low resolution scales, the induced many-body interactions grow very large, possibly being intractable for practical calculations~\citep{Bogner2008AP_InducedForces}.
    Typically one tries to identify the ``right'' resolution window
    (not too low and not too high) where induced many-body interactions that are often neglected remain controllably small.
    \item Crucially, RG transformations to lower resolution also change the physics interpretation.
    Explicit short-range interactions at high resolution
    are supplanted by effective low-energy interactions (often with many-body contributions)~\citep{Tropiano2021PRC_SRCs}.
    Both theoretical models are valid descriptions of the physics, just at different resolutions,
    and the low-resolution description is often simpler.
\end{enumerate}

Let us illustrate some of these concepts using Eq.~\eqref{eq:pt2} as an example again.
We have a Hamiltonian
\begin{equation}
    H = H_0 + H_1^{\Lambda}\,.
\end{equation}
$H_0$ is the unperturbed Hamiltonian. It has eigenstates $\ket{n}$ with eigenvalues $E_n$.
$H_1^{\Lambda}$ is the perturbation, where we are explicit about the resolution scale $\Lambda$.
Up to second order in perturbation theory, our prediction for the ground-state energy is
\begin{align}
    &E^{(0)} = E_0, &E^{(1)}(\Lambda)=\braket{0|H_1^{\Lambda}|0}, &&E^{(2)}(\Lambda) = \sum_{\substack{n\neq0 \\ E_n\leq\Lambda}} \frac{|\braket{0| H_1^{\Lambda}|n}|^2}{E_0 - E_n}\,.
    \label{eq:pt2_full}
\end{align}
We are explicit that the first- and second-order energy corrections $E^{(1)}(\Lambda)$ and $E^{(2)}(\Lambda)$ depend on the resolution scale $\Lambda$
and that the sum over virtual states is already cut off to only include states $\ket{n}$ with $E_n\leq \Lambda$.

Our goal is to determine $H_1^{\Lambda'}$ for $\Lambda' < \Lambda$ such that
\begin{equation}
    \label{eq:RGCondition}
    E^{(1)}(\Lambda') + E^{(2)}(\Lambda') \approx E^{(1)}(\Lambda) + E^{(2)}(\Lambda)\,.
\end{equation}
Note that we do not expect $E^{(1)}(\Lambda') = E^{(1)}(\Lambda)$ and $E^{(2)}(\Lambda') = E^{(2)}(\Lambda)$ separately.
Also, the matching in Eq.~\eqref{eq:RGCondition} does not need to be exact, especially since we are working with a truncated perturbative expansion of the ground state.
Pragmatically, it just needs to be within the precision of the original theory for $H_1^{\Lambda}$ and the theory with which we are solving the problem (in this case second-order perturbation theory).

We write $H_1^{\Lambda'} = H_1^{\Lambda} + \delta H_1$ and consider the case $\Lambda' = \Lambda - \delta\Lambda$ where $\delta\Lambda$ is small.
We may assume $\delta H_1$ is small and neglect terms that are quadratic in $\delta H_1$.
In this case, Eq.~\eqref{eq:RGCondition} simplifies to
\begin{equation}
    \sum_{\substack{n\neq0 \\ \Lambda'<E_n\leq\Lambda}} \frac{|\braket{0| H_1^{\Lambda}|n}|^2}{E_0 - E_n} \approx \braket{0|\delta H_1|0} + \sum_{\substack{n\neq0 \\ E_n\leq\Lambda'}} \frac{\braket{0| H_1^{\Lambda}|n}\braket{n| \delta H_1|0}}{E_0 - E_n}\,.
\end{equation}
To progress further, we would need to be more specific about the Hamiltonian and the RG approach to determine the form of $\delta H_1$.
Still, we clearly see the effects of $H_1^\Lambda$ that we want to remove on the left-hand side.
These are the matrix elements connecting the ground state to states from $\Lambda'$ to $\Lambda$.
These must be effectively captured by new interactions $\delta H_1$ on the right-hand side that only connect the ground state to itself or to states below $\Lambda'$.
This ensures renormalization group invariance of the total ground-state energy $E_\mathrm{gs} \approx E^{(0)} + E^{(1)} + E^{(2)}$.
The resulting $\delta H_1$ contributes at both first and second order in perturbation theory and together these compensate for the decoupled parts of $H_1^\Lambda$.
In particular the nonzero first-order contribution $\braket{0|\delta H_1|0}$ shows that decoupled high-energy physics can be shuffled into low-energy effective operators.
Changing resolution also changes the split of contributions between first, second, and higher orders in perturbation theory.

Additionally, if we suppose $H_1^\Lambda$ and $H_1^{\Lambda'}$ are two-body interactions, we can consider computations of the three-body system using both Hamiltonians.
What one will observe is that $H_1^\Lambda$ and $H_1^{\Lambda'}$ give different three-body binding energies, especially for $\Lambda'$ significantly smaller than $\Lambda$.
This shows that the effective two-body interaction $\delta H_1$ induced through the RG transformation does not on its own guarantee RG invariance of the three-body binding energy
and an induced three-body force is required~\citep{Hammer2013RMP_3BForcesReview}.
Demanding exact RG invariance for the four-body, five-body, and higher-body systems will show that also four-body, five-body, and many-body forces are induced.
How large these are and whether they really need to be accounted for can be studied by looking at the resolution scale dependence of many-body observables, like binding energies, densities, excitation spectra, and so on.

The example in this section is admittedly schematic, but it illustrates many of the key features of the renormalization group, in particular how high-energy physics in the Hamiltonian is decoupled and how it is captured by new effective interactions.
Many renormalization group methods exist, which primarily differ in how the decoupling is performed, how the effective interaction is constructed, and how RG invariance is enforced.
Below we discuss the similarity renormalization group, which has emerged as a popular RG variant in nuclear physics mostly due to its robustness and broad applicability.

\section{The similarity renormalization group}

The similarity renormalization group was first presented by Głazek, Wilson, and Wegner in 1994~\citep{Glazek1994PRD_SRG, Wegner1994AP_SRG} and introduced to nuclear physics over a decade later~\citep{Bogner2007PRC_SRGNN}.
The original publications focused on the renormalization of Hamiltonians in relativistic field theories~\citep{Glazek1994PRD_SRG} and condensed matter systems~\citep{Wegner1994AP_SRG}.
In both cases, strong off-diagonal couplings between low- and high-energy states in the Hamiltonians made them nonperturbative and difficult to solve,
and the common approach developed was a continuous unitary renormalization group transformation to suppress off-diagonal matrix elements and bring the Hamiltonian closer to diagonalization.
We discuss this approach in this section and the application to nuclear forces in the next.

Starting from any Hamiltonian $H$, we want to perform a unitary similarity transformation
\begin{equation}
    H(s) = U(s)\,H\, U^\dagger(s)\,.
    \label{eq:UnitarySimilarityTransform}
\end{equation}
The transformation is a function of a continuous flow parameter $s$,
and $s=0$ is the initial condition with $U(0) = 1$.
Note that this is an operator equation, valid in any basis.
The unitary transformation preserves the eigenvalues of the operator $H$,
leaving the observables intact while changing the nonobservable off-shell behavior.
The transformation of the operator is implicitly also matched by an equivalent transformation of the wave function
such that the result is unchanged.
But, if we do it right,
computing the eigenvalues from the transformed Hamiltonian $H(s)$
will be easier than for the original $H$.

Differentiating Eq.~\eqref{eq:UnitarySimilarityTransform} with respect to $s$ gives the flow equation
\begin{equation}
    \label{eq:SRG}
    \frac{dH(s)}{ds} = [\eta(s), H(s)]
\end{equation}
with the generator $\eta(s)$
and the commutator $[A, B] = AB - BA$.
The anti-Hermitian generator
\begin{equation}
    \eta(s) =\frac{dU(s)}{ds} U^\dagger(s)
\end{equation}
determines the SRG unitary transformation.\footnote{
    Note that $U(s)U^\dagger(s) = 1$,
    which means $dU(s)/ds\,U^\dagger(s) = - U(s)\,dU^\dagger(s)/ds$
    or simply $\eta(s) = - \eta^\dagger(s)$.
}
Rewriting this as
\begin{equation}
   \frac{dU(s)}{ds} =  \eta(s) U(s)
\end{equation}
with the initial condition $U(0)=1$,
one can formally integrate the differential equation
to obtain an expression for the unitary transformation as the $s$-ordered exponential
\begin{align}
    U(s) &= \mathcal{T}_s\left\{\exp \int_0^s  ds'\eta(s')\right\} \nonumber \\
    &= 1 + \int_0^s ds' \eta(s')  + \int_0^s ds'\int_0^{s'} ds'' \eta(s') \eta(s'') + \int_0^s ds'\int_0^{s'} ds'' \int_0^{s''} ds''' \eta(s') \eta(s'') \eta(s''') +\dots\,.
\end{align}
This expression is needlessly cumbersome because one needs to know the generator for all $s$ to perform the integral.
In practice, the SRG unitary transformation is performed by solving the coupled differential equation~\eqref{eq:SRG} from $s=0$ towards $s\to\infty$ instead.
In this way, one only ever needs to store $\eta(s)$, $H(s)$, and $dH(s)/ds$ for the current value of $s$.

The form of the generator $\eta$ is a matter of choice,
and many equivalent choices are possible.
So how do we choose an appropriate generator?
It is instructive to consider the ansatz proposed by Wegner.
We split the Hamiltonian into diagonal and off-diagonal parts\footnote{
    The definition of ``diagonal'' and ``off-diagonal'' parts is arbitrary,
    but one typically chooses a convenient basis
    where the diagonal part is simply the diagonal matrix elements of the Hamiltonian.
}
\begin{equation}
    H = H_d + H_{od}\,.
\end{equation}
We suppress the dependence on the flow parameter $s$, but in general $H_d$ and $H_{od}$ are $s$-dependent.
Wegner suggested the generator
\begin{equation}
    \label{eq:WegnerGenerator}
    \eta = [H_d, H] = [H_d, H_{od}]\,.
\end{equation}
The matrix representation of $H_d$ is just $\varepsilon_i \delta_{ij}$ with the diagonal matrix elements $\varepsilon_i = H_{ii}$.
With this, Eq.~\eqref{eq:WegnerGenerator} simplifies to
\begin{equation}
    \label{eq:WegnerGeneratorMatrixElements}
    \eta_{ij} = H_{ij} (\varepsilon_i - \varepsilon_j)\,.
\end{equation}

We see that the matrix elements $\eta_{ij}$ vanish if the off-diagonal matrix elements $H_{ij}$ ($i\neq j$) vanish.
In this case, the SRG flow equation \eqref{eq:SRG} has reached a fixed point where the decoupling condition,
i.e., the full diagonalization of the Hamiltonian,
is met.\footnote{
    Using the SRG to perform a full diagonalization is neither efficient nor numerically stable
    because the flow equation becomes increasingly stiff as the Hamiltonian reaches a diagonal form.
    In general, the SRG is used to evolve Hamiltonians towards a band-diagonal form,
    where matrix elements outside a band around the diagonal have all been driven to 0.
    Band-diagonal Hamiltonians with sufficiently narrow bands have localized eigenstates, making them easier to solve.
}
The generator matrix elements are proportional to the energy difference $\varepsilon_i - \varepsilon_j$.
As we see below, this is responsible for decoupling states $i$ and $j$ based on their energy differences.

Plugging Eq.~\eqref{eq:WegnerGeneratorMatrixElements} into the flow equation~\eqref{eq:SRG},
we get
\begin{align}
    \frac{dH_{ij}}{ds} &= [\eta, H]_{ij} = -(\varepsilon_i - \varepsilon_j)^2 H_{od,ij} + \sum_{k} (\varepsilon_i + \varepsilon_j - 2 \varepsilon_k)H_{od,ik}H_{od,kj}\,.
\end{align}
To gain more insight into what is going on, we consider on a perturbative analysis where
we assume $H_d \gg H_{od}$.
This is a reasonable starting point,
but of course the degree to which this assumption is met depends on the specifics of the system and problem at hand.
Under this assumption, the first term on the right hand-side proportional to $(\varepsilon_i - \varepsilon_j)^2$ dominates.
This only changes off-diagonal matrix elements,
and if we neglect the second term we can perform the integration to find
\begin{equation}
    H_{ij}(s) = H_{ij}(0)\exp[-(\varepsilon_i - \varepsilon_j)^2 s],\:(i\neq j)\,.
\end{equation}
In this approximation the diagonal matrix elements $\varepsilon_i$ are independent of $s$.
We see that the off-diagonal matrix elements are suppressed as $s$ grows.
How strongly they are suppressed depends on the energy difference $\varepsilon_i - \varepsilon_j$.
From this, we see that we do not need to integrate all the way to $s\to \infty$;
it is good enough to integrate to an $s$
where the problematic large off-diagonal matrix elements have been driven to 0
and the remaining Hamiltonian can be solved.
If we do everything exactly, $H(s)$ at any $s$ is unitarily equivalent to $H(0)$,
giving the same results for observables.

Clearly the generator defined by Wegner is doing what we want,
which is suppressing off-diagonal matrix elements
and driving the Hamiltonian towards the diagonal.
Other choices for the generator are also possible,
and we discuss an alternative that has proliferated in nuclear physics in the next section.
In general, researchers have explored generators of the form
\begin{equation}
    \label{eq:SRGGenerator}
    \eta(s) = [G(s), H(s)]\,,
\end{equation}
where one has to choose the Hermitian matrix $G(s)$.
Picking $G(s)$ to have the form you want to drive the Hamiltonian towards (for example, band diagonal or block diagonal)
has proven to be an effective ansatz to achieve the desired decoupling~\citep{Anderson2008_SRGBlock}.

The SRG induces many-body interactions,
even if the starting Hamiltonian only contains one- and two-body interactions.
To see this we use second quantization to express our Hamiltonian in terms of creation operators $a^\dagger_i$ and annihilation operators $a_i$ (with $a^\dagger_i\ket{0}=\ket{i}$):
\begin{equation}
    H = T + V = \sum_{ij} t_{ij}\, a^\dagger_i a_j + \frac{1}{4}\sum_{ijkl} v_{ijkl}\, a^\dagger_i a^\dagger_j a_l a_k\,.
\end{equation}
The one-body operator $T$ (for example, the kinetic energy or an external potential)
is expressed in terms of its matrix elements $t_{ij}$, which parametrizes interactions between particles in the single-particle states $\ket{i}$ and $\ket{j}$.
Similarly, the two-body operator $V$ with matrix elements $v_{ijkl}$ generates interactions between pairs of particles,
for example, taking particles from the states $\ket{k},\ket{l}$ to $\ket{i},\ket{j}$.
The generator also has one- and two-body parts:
\begin{equation}
    \eta = \sum_{ij} \eta_{ij}\, a^\dagger_i a_j + \frac{1}{4}\sum_{ijkl} \eta_{ijkl}\, a^\dagger_i a^\dagger_j a_l a_k\,.
\end{equation}
Evaluating the commutator in Eq.~\eqref{eq:SRG} gives expressions of the form
\begin{equation}
    \label{eq:SecondQuantizedCommutator}
    \big[a^\dagger_i a^\dagger_j a_l a_k, a^\dagger_a a^\dagger_b a_d a_c\big] = 
    a^\dagger_i a^\dagger_j a_l a_k a^\dagger_a a^\dagger_b a_d a_c - a^\dagger_a a^\dagger_b a_d a_c a^\dagger_i a^\dagger_j a_l a_k\,.
\end{equation}
Each of the terms on the right-hand side must be simplified using normal ordering techniques,
and after some work one finds that there are terms proportional to
$a^\dagger_p a^\dagger_q a^\dagger_r a_u a_t a_s$ and $a^\dagger_p a^\dagger_q a^\dagger_r  a^\dagger_s a_w a_v a_u a_t$,
which are three- and four-body operators, respectively.
The subtraction in Eq.~\eqref{eq:SecondQuantizedCommutator} causes the four-body operators to cancel exactly,
but the three-body operators remain.
These are induced three-body interactions generated by the SRG transformation, and further iteration introduces many-body forces.

If we are only interested in two-body systems, three-body interactions do not contribute
so we do not need to care and can solve the SRG equations in the two-body Hilbert space.
However, if we look at three-body systems, 
for the SRG transformation to be perfectly unitary,
which means leaving all eigenvalues and observables unchanged,
we need to solve the SRG equations in the three-body Hilbert space.
Doing this allows us to capture the induced three-body interactions,
which are essential to guarantee exact renormalization group invariance as $s$ goes from 0 towards $\infty$.
Four-body interactions are also induced,
and capturing these would require evaluating the SRG equations in the four-body Hilbert space.
Solving the SRG equations in the four-body Hilbert space and beyond is currently impractical,
so induced many-body interactions beyond three-body forces are not accounted for.
The size of these interactions can be probed by checking RG invariance:
How much does a prediction for an observable depend on the value of $s$?
In general, one evaluates the SRG transformation to a finite value of $s$.
RG invariance is then tested by varying $s$ in some range.
Ideally, we find that varying $s$ only changes predicted observables a little,\footnote{
How much or little ultimately depends on the precision of our theory.
If we are using a leading order effective theory with uncertainties of $\sim$30\,\%,
it may be acceptable to have significant, but controlled dependence on $s$ within this uncertainty.
}
which indicates that missing induced many-body interactions are small.
If we take $s$ very large, however, we will see significant dependence on $s$,
indicating that neglecting induced many-body interactions is no longer a controlled approximation.

\section{Generating low-resolution nuclear forces}

Nuclear forces have been traditionally thought of as nonperturbative,
with a hard repulsive core required to give nuclear saturation~\citep{Bethe1971ARNPS_NuclearMatterReview, Lacombe1980PRC_ParisNN, Wiringa1995PRC_AV18, Machleidt2001PRC_CDBonn} (see also the chapter~\cite{Elster_ENP}, ``The nuclear force: the first sixty years'' in this volume).
This hard core strongly couples low- and high-energy states,
preventing any perturbative description of nuclei.
The renormalization group tells us that this picture is more complicated than necessary.
Such hard core potentials are one possible paradigm for nuclear forces,
but other lower-resolution theories and models are also possible.
In particular, modern potentials from chiral and pionless effective field theory may be constructed at lower resolution,
making them more perturbative and easier to solve in many-body calculations.
The SRG allows us to further transform nuclear Hamiltonians
(from effective field theory, phenomenological fits, or other methods)
to even lower resolution,
making them even more perturbative.

Nuclear Hamiltonians generally have the form
\begin{equation}
    H = T - T_\mathrm{cm} + V_\mathrm{NN} + V_\mathrm{3N}\,.
\end{equation}
We have the intrinsic kinetic energy $T - T_\mathrm{cm}$
with the center of mass subtracted
(often also called $T_\mathrm{int}$ or $T_\mathrm{rel}$),
a nucleon-nucleon potential $V_\mathrm{NN}$,
and a three-nucleon potential $V_\mathrm{3N}$.
We are neglecting four-nucleon interactions,
which exist and are for example predicted at next-to-next-to-next-to-leading order in chiral effective field theory~\citep{Epelbaum2009RMP_ChiralEFTReview, Machleidt2011PR_ChiralEFTReview}
(see also the chapters~\cite{Hammer_ENP}, ``Renormalization and effective field theory'' and~\cite{Epelbaum_ENP}, ``Chiral effective field theory for nuclear forces'' in this volume).

In this section, we focus on the SRG for nucleon-nucleon forces
and only briefly discuss the application to three-nucleon forces~\citep{Jurgenson2009PRL_SRG_ManyBody, Hebeler2012PRC_SRG3N}.
Nucleon-nucleon forces may be represented in a relative-momentum plane-wave partial-wave basis
\begin{equation}
    \ket{p (L S)J M_J T M_T} \equiv \ket{p\, \alpha_\mathrm{NN}}.
\end{equation}
$p = |\mathbf{p}|$ is the magnitude of the relative momentum $\mathbf{p} = (\mathbf{k}_1 - \mathbf{k}_2) / 2$,
where $\mathbf{k}_1$ and $\mathbf{k}_2$ are the momenta of the first and second nucleon, respectively.
The isospin and angular momentum dependence of the potential is captured in the partial-wave quantum numbers $\alpha_\mathrm{NN}$:
the relative orbital angular momentum $L$,
the total two-body spin $S$,
the coupled angular momentum $J$ with its projection $M_J$,
and the two-body isospin $T$ with its projection $M_T$.
For simplicity, we focus on singlet channels, where $S=0$, where the potential is diagonal in all partial-wave quantum numbers.\footnote{
Symmetries of nuclear forces require that all nucleon-nucleon partial-wave quantum numbers are diagonal
except $L$, which is $J\pm1$ in coupled channels where $S=1$.
To generalize the expressions for singlet channels,
replace integrals over intermediate momenta
\begin{equation}
    \frac{2}{\pi}\int q^2 dq
\end{equation}
with an additional sum over possible intermediate channels
\begin{equation}
    \frac{2}{\pi} \sum_{\alpha} \int q^2 dq\,.
\end{equation}
}
In this case,
we can write nucleon-nucleon potentials simply as
\begin{equation}
    V(p', p) = \braket{p' \alpha_\mathrm{NN}| V_\mathrm{NN} | p\,\alpha_\mathrm{NN}}.
\end{equation}

In nuclear physics, choosing the generator as
\begin{equation}
    \eta(s) = [T_\mathrm{rel}, H(s)]
\end{equation}
has emerged as a standard choice~\citep{Glazek2008PRD_SRG, Wendt2011PRC_SpuriousSRG}.
This follows the established intuition for the SRG
that,
since $T_\mathrm{rel}$ is diagonal in the plane-wave partial-wave basis,
choosing $G=T_\mathrm{rel}$ in Eq.~\eqref{eq:SRGGenerator}
will drive $H(s)$ towards the diagonal.
Plugging all of this into the momentum-space SRG flow equation
gives:\footnote{
We use the normalization
\begin{equation}
    1 = \frac{2}{\pi}\int_0^\infty q^2\, dq \, \ket{q}\bra{q}.
\end{equation}
We use units such that $\hbar c = 1$ and the nucleon mass $M_\mathrm{N} = 1$.
$T_\mathrm{rel}$ then has matrix elements $T_\mathrm{rel}(p', p) = p^2 \frac{\pi}{2}\frac{\delta(p - p')}{pp'}$.
}
\begin{equation}
    \label{eq:SRGNucleonNucleon}
    \frac{dV_s(p',p)}{ds} = - (p'^2 - p^2)^2 V_s(p',p) + \frac{2}{\pi}\int_0^\infty q^2\,dq\,(p^2 + p'^2 - 2 q^2) V_s(p, q) V_s(q, p')\,.
\end{equation}

Assuming $T_\mathrm{rel} \gg V_s$, we see that the first term dominates, giving
\begin{equation}
    V_s(p',p) = V_{s=0}(p',p) \exp[-(p'^2 - p^2)^2 s]\,.
\end{equation}
off-diagonal matrix elements are suppressed based on the kinetic energy difference $p'^2 - p^2$.
Changing variables to
\begin{equation}
    \lambda = \frac{1}{s^{1/4}}\,,
\end{equation}
where $\lambda$ has units of \fmi{},
helps us to further analyze the behavior of the SRG equation.
Now the SRG equations are instead integrated from $\lambda=\infty$ towards $\lambda = 0$,
and
\begin{equation}
    \label{eq:SRGSuppression}
    V_\lambda(p',p) = V_{\infty}(p',p) \exp[-(p'^2 - p^2)^2 / \lambda^4]\,.
\end{equation}
This change of variables is useful because $\lambda$ has the clear interpretation of a resolution scale
that is lowered by the SRG transformation as we integrate towards $\lambda=0\:\fmi$,
and off-diagonal matrix elements outside of a band of width $\lambda^2$ are exponentially suppressed.

For nucleon-nucleon potentials, solving the SRG is simple.
Equation~\eqref{eq:SRGNucleonNucleon} can be solved on a laptop using established adaptive differential equation solvers.
Accompanying this chapter is a Python code \texttt{\detokenize{nn-srg}}~\citep{Heinz2025_NNSRGGit, Heinz2025_NNSRGZenodo} to allow readers to experiment with the SRG transformation of nucleon-nucleon potentials.
Let us now turn our attention to the SRG in practice. 

\subsection*{Low-resolution nucleon-nucleon potentials}

\begin{figure}[t!]
    \centering
    \includegraphics{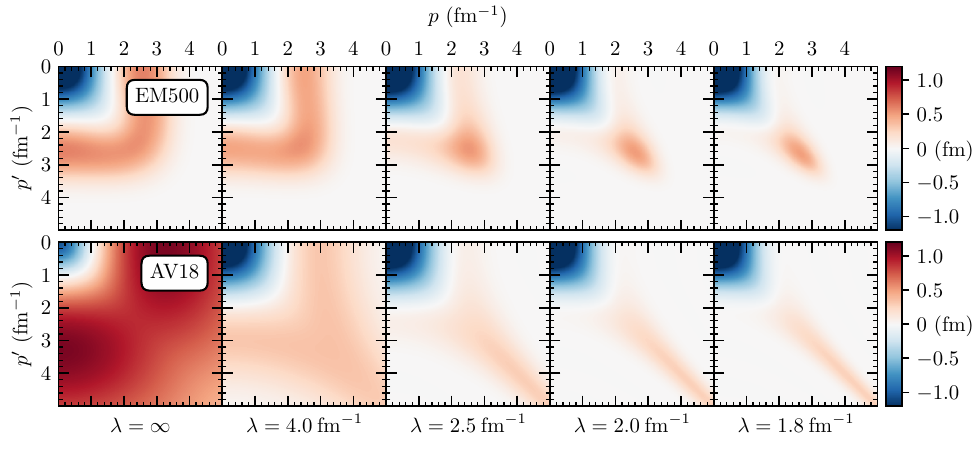}
    \caption{Relative momentum-space nucleon-nucleon potential matrix elements $V(p',p)$ in the neutron-proton $^1\mathrm{S}_0$ channel for various Hamiltonians.
    We consider two initial Hamiltonians: the ``EM500'' nucleon-nucleon potential (top) from chiral effective field theory at next-to-next-to-next-to-leading order~\citep{Entem2003PRC_EM500};
    and the phenomenological ``AV18'' nucleon-nucleon potential (bottom)~\citep{Wiringa1995PRC_AV18}.
    These potentials are constructed very differently.
    AV18 is obtained from a largely phenomenological parametrization that is almost entirely local (meaning the potential is just a function of the relative distance $r$ between two nucleons),
    which leads to a repulsive short-range core for small $r$.
    Due to its repulsive short-range core, AV18 clearly couples low and high momenta strongly
    with nonzero matrix elements $V(0, p)$ beyond $p=5\:\fmi$.
    On the other hand, EM500 potential is constructed using a separable nonlocal regulator with a momentum-space cutoff $\Lambda$.
    We clearly see that the $\Lambda=500\:\mev$ cutoff of the EM500 potential regularizes high-momentum behavior such that matrix elements with $p, p' \gtrsim 3.2\:\fmi$ are approximately 0.
    For both potentials, we show the SRG transformation to lower resolution scales $\lambda$, down to $\lambda=1.8\:\fmi$.
    We see that the potentials are driven towards a band-diagonal form, and matrix elements $V(0, p)$ with $p>\lambda$ are suppressed.
    Once SRG transformed to low resolution, AV18 and EM500 have very similar low-momentum matrix elements.}
    \label{fig:SRGContours}
\end{figure}

\begin{figure}[t!]
    \centering
    \includegraphics{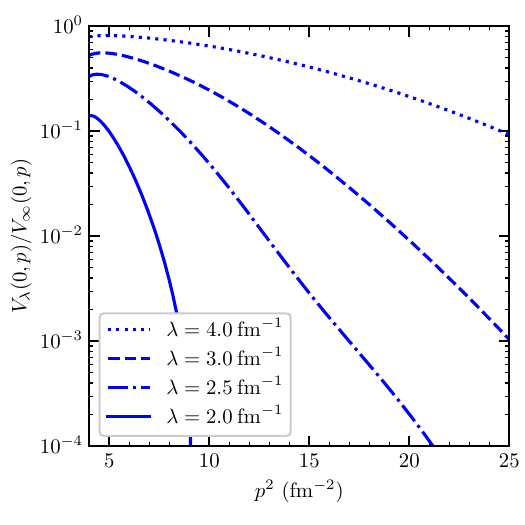}
    \caption{Matrix element suppression for the matrix elements $V_\lambda(0, p)$ visualized through the ratio $V_\lambda(0, p)/ V_\infty(0, p)$.
    The initial Hamiltonian is the EM500 potential in the neutron-proton $^1\mathrm{S}_0$ channel.
    As the EM500 potential is SRG transformed to lower resolution scales $\lambda$, the off-diagonal matrix elements with $p>\lambda$ are exponentially suppressed
    as predicted in Eq.~\eqref{eq:SRGSuppression}.}
    \label{fig:SRGSuppression}
\end{figure}

In Fig.~\ref{fig:SRGContours},
we visualize the evolution of nuclear potentials as they are transformed to lower resolution using the SRG.
Each panel shows $V(p',p)$ in the neutron-proton $^1\mathrm{S}_0$ channel with $p',p$ up to $5\:\fmi$.
The leftmost panel is the untransformed potential (as originally constructed from the effective field theory or phenomenological methods).
Going to the right, we transform the potential using the SRG to lower resolution scales $\lambda$ all the way down to $\lambda=1.8\:\fmi$.
The top row considers the EM500 potential~\citep{Entem2003PRC_EM500}, constructed at next-to-next-to-next-to-leading order in chiral effective field theory.
Looking at the initial potential on the left, we see that matrix elements with $p',p\gtrsim 3.2\:\fmi$ are approximately 0.
This is due to the $\Lambda=500\:\mev\approx2.5\:\fmi$ cutoff used in the construction of the EM500 potential, which sets its resolution scale.
The SRG transformation to lower resolutions initially does very little.
$\lambda = 4.0\:\fmi$ and $\lambda=\infty$ are nearly identical.
Once $\lambda \leq \Lambda$ the off-diagonal matrix elements begin to be suppressed, and the SRG transformation drives the potential towards a band-diagonal form as desired.
At $\lambda=1.8\:\fmi$, all matrix elements $V(0,p)$ with $p>1.8\:\fmi$ are strongly suppressed,
so low- and high-momentum degrees of freedom have been successfully decoupled.
In Fig.~\ref{fig:SRGSuppression}, we see that this suppression of off-diagonal matrix elements is exponential
as predicted by Eq.~\eqref{eq:SRGSuppression}.

For comparison, we also show a phenomenological nucleon-nucleon potential, AV18, in the bottom row of Fig.~\ref{fig:SRGContours}~\citep{Wiringa1995PRC_AV18}.
AV18 has a hard repulsive core,
which leads to strong coupling between low and high momenta.
Clearly the matrix elements $V_\infty(0,p)$ are sizable up to and beyond $p=5\:\fmi$.
The AV18 potential has an intrinsically high resolution.
This is due to its construction, particularly the requirement of locality.
Nonetheless, the SRG transformation systematically drives the potential to a diagonal form,
successfully decoupling low and high momenta in the low-resolution potential.
At $\lambda=1.8\:\fmi$, the low-resolution potentials obtained from the EM500 and AV18 potentials at $\lambda=\infty$ have very similar low-momentum matrix elements.
Both potentials share the same long-range physics due to one-pion exchange
and are optimized to reproduce nucleon-nucleon scattering data,
so the model-dependent choices for how to handle short-range physics
``collapse'' to universal forms at low resolution~\citep{Bogner2003PLB_Vlowk, Bogner2003PR_VlowkReview, Bogner2010PPNP_RGReview, Dainton2014PRC_SRGUniversality}.

\subsection*{Induced many-body forces}

\begin{figure}[t!]
    \centering
    \includegraphics[width=0.9\linewidth]{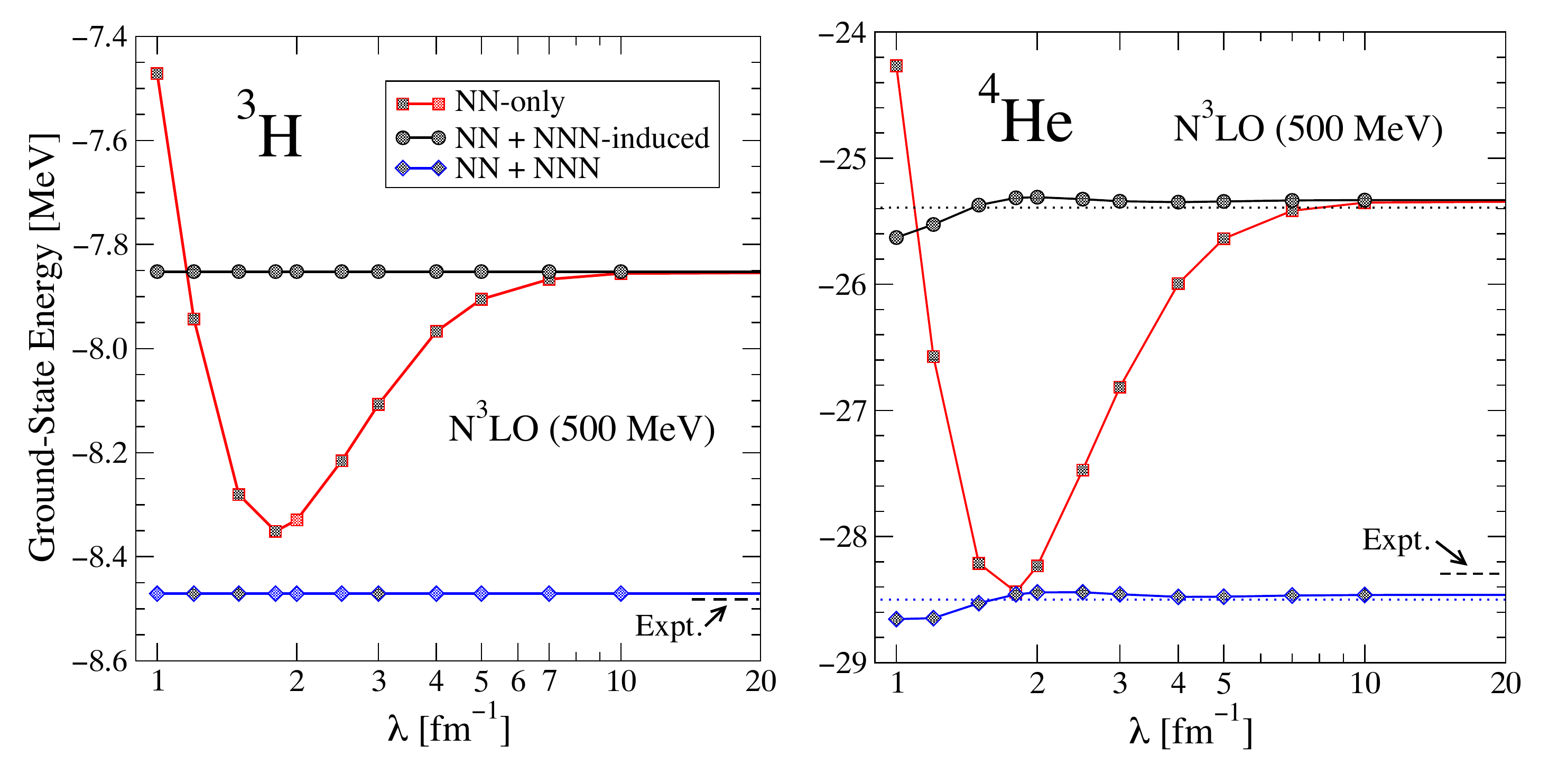}
    \caption{The ground-state energies of $^3$H (left) and $^4$He (right) computed with nuclear Hamiltonians from chiral effective field theory
    that have been transformed using the similarity renormalization group to lower resolution scales $\lambda$.
    We compare the behavior of the ground-state energy for various $\lambda$ when only SRG-evolved nucleon-nucleon potentials are used (NN-only, red line),
    when the SRG-induced three-nucleon interactions are included (NN~+~NNN-induced, black line),
    and when a three-nucleon potential is included in the initial Hamiltonian (NN~+~NNN, blue line).
    For $^3$H, accounting for induced three-nucleon interactions is required for SRG invariance of the ground-state energy,
    and the inclusion of explicit three-nucleon forces is required to reproduce the experimental value.
    For $\lambda \geq 1.8\:\fmi$,
    the predicted ground-state energy of $^4$He is approximately independent of $\lambda$,
    indicating that contribution of induced four-body forces is small.
    Figure adapted from~\cite{Jurgenson2009PRL_SRG_ManyBody}.}
    \label{fig:TritonFlow}
\end{figure}

We now clearly see how the SRG transforms nuclear forces to lower resolution.
This transformation is unitary,
and it leaves nucleon-nucleon observables and nucleon-nucleon phase shifts unchanged
while decoupling low- and high-momentum degrees of freedom.
As we know, RG transformations induce many-body interactions,
and the SRG is no exception.
In the left panel of Fig.~\ref{fig:TritonFlow},
we see this for the $^3$H ground-state energy.
When only the nucleon-nucleon interactions are transformed using the SRG (NN-only),
the ground-state energy is not invariant as a function of $\lambda$.
This is due to missing induced three-body interactions.

Consistently performing the SRG transformation for two- and three-nucleon forces is challenging but possible.
In general, one needs to perform the SRG transformation for the two- and three-body Hamiltonians at the same time:
\begin{align}
    H_2(s) &= U(s) (T_\mathrm{rel} + V_\mathrm{NN})U^\dagger(s)\,, \\
    H_3(s) &= U(s) (T_\mathrm{rel} + V_\mathrm{NN} + V_\mathrm{3N})U^\dagger(s)\,.
\end{align}
At every $s$, one can unambiguously identify the new nucleon-nucleon and three-nucleon potentials sequentially:
\begin{align}
    V_\mathrm{NN}(s) &= H_2(s) - T_\mathrm{rel}\,, \\
    V_\mathrm{3N}(s) &= H_3(s) -T_\mathrm{rel} - V_\mathrm{NN}(s)\,. 
\end{align}
If we include no initial three-nucleon forces
[i.e., $V_\mathrm{3N}(s=0) = 0$],
the induced three-nucleon potential $V_\mathrm{3N}(s)$ ensures that, for example, the $^3$H ground-state energy is invariant under the SRG transformation (see NN~+~NNN-induced in Fig.~\ref{fig:TritonFlow}).
Three-nucleon forces are always present and are essential for an accurate description of nuclei and nuclear matter~\citep{Hammer2013RMP_3BForcesReview, Hebeler2015ARNPS_3NFNeutronRichReview, Hebeler2021PR_3NF}.
Including the initial three-nucleon potential in the Hamiltonian in Fig.~\ref{fig:TritonFlow} brings the $^3$H ground-state energy into agreement with experiment at $\lambda=\infty$ (see NN~+~NNN).
Moreover, the SRG transformation does not change $E_\mathrm{gs}$,
indicating that $V_\mathrm{NN}(s)$ and $V_\mathrm{3N}(s)$ are being transformed consistently to ensure that the full three-body Hamiltonian is unitarily equivalent to the initial Hamiltonian.

Performing the SRG consistently for two- and three-nucleon potentials is the current state-of-the-art in nuclear physics~\citep{Jurgenson2009PRL_SRG_ManyBody, Roth2011PRL_SRG3N, Hebeler2012PRC_SRG3N, Hebeler2021PR_3NF}.
Both initial and induced four-nucleon forces are neglected.
This is not necessarily a problem
as the contributions of four-nucleon interactions in nuclei may very well be smaller than other uncertainties in our calculations.
Still this uncertainty should be studied and quantified, and we currently rely on variations of the SRG resolution scale $\lambda$ as a diagnostic of the uncertainty associated with neglected four-body and many-body forces.
As one example, we see the impact of missing induced four-nucleon forces in calculations of the ground-state energy of $^4$He in the right panel of Fig.~\ref{fig:TritonFlow}.
Even with consistent SRG transformation of two- and three-nucleon potentials, the ground-state energy varies weakly as $\lambda$ is reduced below $3\:\fmi$
and more strongly as $\lambda$ approaches $1\:\fmi$.
The effect of the induced four-nucleon forces is small overall, and we can conclude that for calculations of $^4$He neglecting these induced forces is a small, controlled approximation.

The SRG is a mature method in nuclear physics, but induced many-body interactions are still a significant challenge.
As we see in the next section,
low-resolution forces make calculations of nuclei much easier, reducing computational costs by orders of magnitude.
This is especially true for very soft interactions,
with $\lambda \lesssim 2.0\:\fmi$.
However, SRG transformations to such low resolution carry the risk of inducing significant four-nucleon interactions.
Quantifying the size of these is difficult, so we must tread with caution.
Any improvement to how many-body forces are induced in the SRG
(for example, by developing a generator that induces smaller many-body forces or by approximately accounting for induced four-nucleon forces via low-resolution fits~\citep{Hebeler2011PRC_SRG3NFits})
would have a significant positive impact on calculations of nuclei.

\section{Implications for nuclei}

Solving the many-body Schrödinger equation is a challenging task,
and for a long time first-principles (or ab initio) calculations based on nuclear forces were limited to light nuclei~\citep{Hergert2020FP_AbInitioReview}
(see the chapter~\cite{Stroberg:2026feo}, ``What we talk about when we talk about nuclear structure'' in this volume for details).
Now comprehensive studies of medium-mass nuclei~\citep{Hagen2012PRL_CaEOMCC, Soma2014PRC_SCGFChains, Hergert2013PRL_MRIMSRG, Hagen2016NP_Ca48Skin, Hagen2016PRL_Ni78, Stroberg2017PRL_VSIMSRG, Stroberg2021PRL_AbInitioLimits, Fearick2023PRR_Ca40AlphaD, Sun2025PRX_Multiscale, Heinz2025PRC_Calcium, Heinz2025PLB_MuToE, Plies2025arxiv_LowResolutionUQ, He:2026zie} and also leadership calculations of heavy~\citep{Miyagi2022PRC_NO2B, Hu2022NP_Pb208, Hebeler2023PRC_JacobiNO2B, Door2025PRL_YbBoson, Miyagi2025PLB_Rch4, Reed2025PRC_RMFChiralEFT, Arthuis2024arxiv_LowResForces, Noel:2026byc} and even super-heavy isotopes~\citep{Bonaiti2025arxiv_Pb266} are possible.
This advance was driven by two combined efforts:
1) the development of low-resolution nuclear Hamiltonians using effective field theories and renormalization group methods~\citep{Epelbaum2009RMP_ChiralEFTReview, Bogner2010PPNP_RGReview, Machleidt2011PR_ChiralEFTReview, Hammer2020RMP_NuclearEFT};
and 2) the development of scalable many-body methods that solve the Schrödinger equation in an approximate, but systematically improvable way~\citep{Hagen2014RPP_CCReview, Hergert2016PR_IMSRG, Tichai2020FP_MBPTReview, Soma2020FP_SCGFReview}.
The SRG was a key method in the development of low-resolution Hamiltonians and is now part of the foundation of modern nuclear theory.

The essential role of low-resolution forces in nuclear structure theory cannot be overstated. 
Many methods to solve the Schrödinger equation rely on basis expansions,
which expand nuclear Hamiltonians in a basis of harmonic oscillator wave functions.
Their computational costs scale with the size of the basis $N$,
either exponentially $\mathcal{O}(e^N)$ for exact diagonalizations
or polynomially, e.g., $\mathcal{O}(N^6)$, for approximate, scalable many-body methods.
Low-resolution Hamiltonians have significantly reduced couplings to high-energy states.
This accelerates convergence in the basis truncation,
ultimately allowing for converged calculations in smaller bases.
This is the main benefit of low-resolution forces in calculations of nuclei.

\begin{figure}[t!]
    \centering
    \includegraphics[trim={4cm 0.8cm 3.6cm 0.5cm},clip,width=0.5\linewidth]{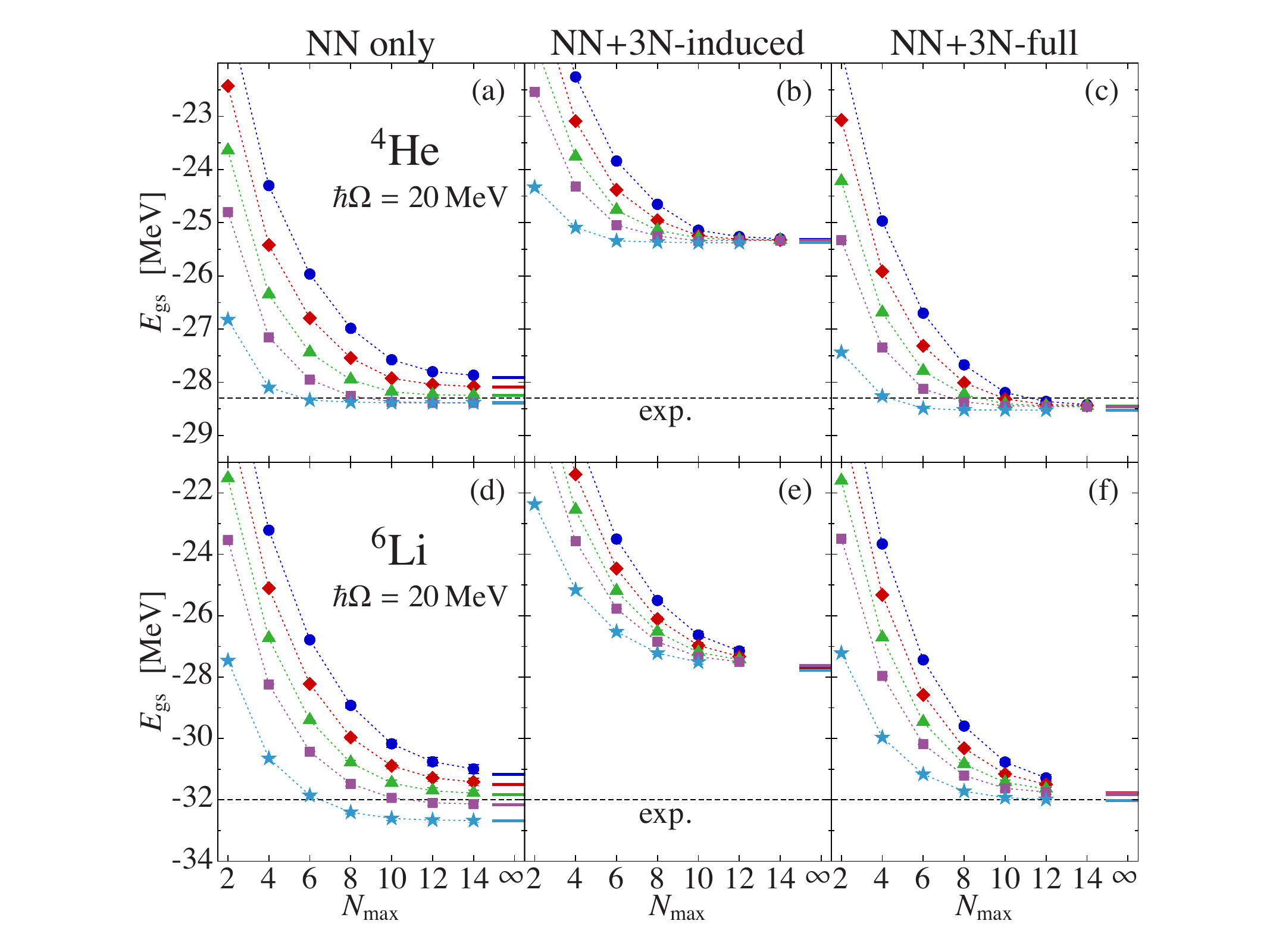}
    \caption{
    Ground-state energies of $^4$He (top)
    and $^6$Li (bottom) computed with the no-core shell model
    using Hamiltonians from chiral effective field theory
    that have been transformed to lower resolution scales $\lambda$.
    The SRG scales are $\lambda = 2.2\:\fmi$ (dark blue circles), $2.1\:\fmi$ (red diamonds), $2.0\:\fmi$ (green triangles), $1.9\:\fmi$ (purple squares), and $1.6\:\fmi$ (teal stars).
    In no-core shell model calculations, the Hamiltonian is expanded in a basis truncated based on the parameter $N_\mathrm{max}$.
    The ground-state energies are shown as a function of $N_\mathrm{max}$,
    where the extrapolation to $N_\mathrm{max}\to\infty$ is indicated on the very right.
    As in Fig.~\ref{fig:TritonFlow},
    we consider cases where only the nucleon-nucleon force has been SRG transformed (NN only),
    where the SRG-induced three-nucleon interactions are accounted for (NN+3N-induced),
    and where the Hamiltonian also includes an initial three-nucleon force before the SRG transformation (NN+3N-full).
    As in Fig.~\ref{fig:TritonFlow}, consistently transforming both two- and three-nucleon potentials is required to achieve SRG invariance
    and the inclusion of the the initial three-nucleon force is important for agreement with experiment.
    Comparing $\lambda = 2.2\:\fmi$ and $1.6\:\fmi$ (dark blue circles and teal stars, respectively),
    the ground-state energy converges much more quickly in $N_\mathrm{max}$ for $\lambda=1.6\:\fmi$.
    This improved convergence substantially reduces the computational cost required to deliver converged calculations
    and makes calculations in heavier nuclei up to around $^{16}$O possible.
    Figure from~\cite{Roth2011PRL_SRG3N}.}
    \label{fig:NCSMConvergence}
\end{figure}

\begin{figure}[t!]
    \centering
    \includegraphics[width=0.65\linewidth]{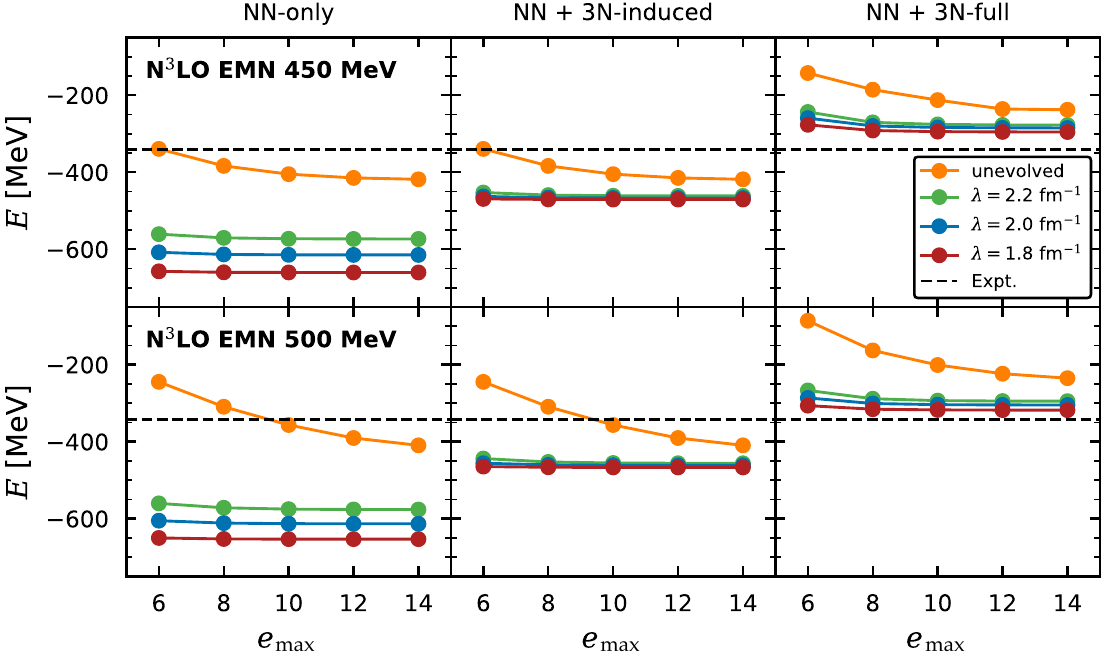}
    \caption{Ground-state energies of $^{40}$Ca computed with the in-medium similarity renormalization group using Hamiltonians from chiral effective field theory
    that have been transformed to lower resolution scales $\lambda$.
    The SRG scales are unevolved, i.e., $\lambda = \infty$, (orange), $2.2\:\fmi$ (green), $2.0\:\fmi$ (blue), and $1.8\:\fmi$ (red).
    In in-medium similarity renormalization group calculations, the Hamiltonian is expanded in a single-particle basis truncated based on the parameter $e_\mathrm{max}$.
    As in Fig.~\ref{fig:TritonFlow},
    we consider cases where only the nucleon-nucleon force has been SRG transformed (NN-only),
    where the SRG-induced three-nucleon interactions are accounted for (NN~+~3N-induced),
    and where the Hamiltonian also includes an initial three-nucleon force before the SRG transformation (NN~+~3N-full).
    Unevolved Hamiltonians converge slowly in $e_\mathrm{max}$ in all cases,
    while SRG-transformed Hamiltonians converge quickly.
    This improved convergence substantially reduces the computational cost required to deliver converged calculations.
    Figure adapted from~\cite{Hoppe2019PRC_ChiralMedMass}.
    }
    \label{fig:IMSRGConvergence}
\end{figure}

Let us explore two examples of improved convergence in calculations of nuclei.
In Fig.~\ref{fig:NCSMConvergence},
we show calculations of the ground-state energies of $^4$He and $^6$Li using the no-core shell model~\citep{Roth2011PRL_SRG3N}.
The no-core shell model performs an exact diagonalization of the many-body Hamiltonian,
where the Hamiltonian is expanded in a basis of $A$-body states (see the chapter~\cite{McCoy_ENP}, ``Large-basis diagonalization for nuclear structure: the no-core shell model \& extensions'' in this volume for details).
This basis is truncated based on $N_\mathrm{max}$,
limiting the maximum number of excitations in a state relative to the lowest energy state (with $N_\mathrm{max}=0$).
No-core shell model calculations must take $N_\mathrm{max}$ large enough to observe that their calculations become independent of this truncation
at which point the calculation is converged.
The cost of the diagonalization scales exponentially in $N_\mathrm{max}$,
so reducing the $N_\mathrm{max}$ required to reach convergence can easily lower the cost of a calculation by orders of magnitude.

Tying this back to the discussion of resolution scales around Eqs.~\eqref{eq:pt2} and~\eqref{eq:pt2_full} previously,
an exact no-core shell model calculation must be performed in an infinite basis with $N =0$ up to $\infty$.
This is like the unrestricted sum over states in Eq.~\eqref{eq:pt2}.
In practice, the Hamiltonian has a given resolution,
which limits the degree of coupling between low-lying states with $N =0$ and high-lying states with $N\gg 0$.
This means that only a finite, but potentially large number of states with $N\leq N_\mathrm{max}$ is required to give quasi exact results for the ground state energy.
Transforming the Hamiltonian to lower resolution with the SRG will further decrease the coupling between low- and high-lying states,
reducing the $N_\mathrm{max}$ required to give exact results.
This is what we see in Fig.~\ref{fig:NCSMConvergence}.

First, consider calculations only using nucleon-nucleon interactions (left column, NN only).
The different colored points correspond to different SRG resolution scales $\lambda$.
The dark blue circles are $\lambda=2.2\:\fmi$,
while the teal stars are $\lambda=1.6\:\fmi$.
For $\lambda=2.2\:\fmi$,
the ground-state energy of $^4$He is only converged at $N_\mathrm{max}=12$,
and that of $^6$Li is not converged even at $N_\mathrm{max}=14$ and requires extrapolation to $N_\mathrm{max}\to\infty$.
We contrast this with $\lambda=1.6\:\fmi$,
where converged ground-state energies are obtained for $^4$He and $^6$Li at $N_\mathrm{max}=6$ and $8$, respectively.

One thing to note here is that the converged results for different lines corresponding to different $\lambda$ do not agree.
This is because induced three-nucleon interactions are not accounted for.
In the center and right columns, we see that including induced and initial three-nucleon interactions gives converged predictions that agree between all $\lambda$.
We also see that the nuclear forces SRG transformed to $\lambda=1.6\:\fmi$ still converge much more quickly in $N_\mathrm{max}$.
The ground-state energy of $^6$Li is converged at $N_\mathrm{max}=8$,
which is 100--1000 times cheaper to compute than $N_\mathrm{max}=12$,
which is still not sufficient to provide a converged result for $\lambda=2.2\:\fmi$.
The improved convergence and corresponding reduction in cost
extends the reach of the no-core shell model,
enabling calculations of systems that could not be computed otherwise.
However, if we look closely, we see that for $^6$Li in the NN+3N-full case, the result for $\lambda=1.6\:\fmi$
does not fully agree with the results for other $\lambda$.
This difference is due to neglected induced four-nucleon forces,
which grow larger when we reach very low resolution scales like $\lambda=1.6\:\fmi$.

In Fig.~\ref{fig:IMSRGConvergence},
we consider instead ground-state energies of $^{40}$Ca
computed using the in-medium similarity renormalization group (IMSRG)~\citep{Hoppe2019PRC_ChiralMedMass}.
The IMSRG~\citep{Tsukiyama2011PRL_IMSRG, Hergert2016PR_IMSRG} is a many-body method based on the SRG idea of decoupling, but applied to the quantum many-body problem
(see the chapter~\cite{Fossez_ENP}, ``The in-medium similarity renormalization group \& coupled cluster methods'' in this volume for details).
The IMSRG differs from the SRG in its decoupling strategy; 
it performs a block diagonalization of the many-body Hamiltonian, starting from a reference state that approximates the ground state of the nucleus of interest
and decoupling only that reference state from its excitations.
Thus the IMSRG transformation is specific to each system computed (hence the name ``in-medium'' SRG),
which allows the IMSRG to efficiently solve the many-body problem.
However, this also prevents it from being used to derive general low-resolution Hamiltonians
like the ``free-space'' SRG discussed in this chapter.

In IMSRG calculations,
the Hamiltonian is expanded in a basis of single-particle harmonic oscillator states.
This basis is truncated,
including only states with a harmonic oscillator energy up to a given value $e_\mathrm{max}$.
As with the no-core shell model,
IMSRG calculations must show that their predictions are sufficiently converged with respect to $e_\mathrm{max}$.
The IMSRG scales polynomially in the basis size,
so calculations with larger $e_\mathrm{max}$ are not as expensive as similar increases in $N_\mathrm{max}$.
Still, a calculation with $e_\mathrm{max}=10$ is approximately 20--30 times cheaper than a calculation with $e_\mathrm{max}=14$,
making the improved convergence of SRG-transformed Hamiltonians very valuable.

The trends in Fig.~\ref{fig:IMSRGConvergence} are very similar to those in Fig.~\ref{fig:NCSMConvergence}.
The SRG-unevolved Hamiltonians (orange) converge slowly in $e_\mathrm{max}$,
especially when three-nucleon forces are included.
The Hamiltonians that have been SRG evolved to lower resolution scales $\lambda$ (green, blue, and red) all exhibit much faster convergence with respect to $e_\mathrm{max}$.
$e_\mathrm{max}=10$ is converged in all cases,
and when only nucleon-nucleon forces are included
even $e_\mathrm{max}=8$ is converged.
The results with only nucleon-nucleon forces are of course strongly dependent on $\lambda$ because induced three-body forces have not been accounted for.
Analogous findings also apply to other many-body methods,
such as many-body perturbation theory~\citep{Tichai2020FP_MBPTReview}, coupled cluster theory~\citep{Hagen2014RPP_CCReview}, and self-consistent Green's function theory~\citep{Soma2020FP_SCGFReview},
when they are expanded in harmonic oscillator bases.

\section{Variations on low-resolution forces}

When low-resolution nuclear forces were introduced,
it did not take long for them to lead to many successful applications in calculations of nuclei~\citep{Hagen2007PRC_CCBenchmark, Bogner2008NPA_SRGNCSMConvergence, Quaglioni2008PRL_Scattering}.
Over the past 20 years,
low-resolution forces have remained very successful,
and different approaches to constructing them have proliferated.

Before the SRG was introduced in nuclear physics, $V_{\mathrm{low}\:k}$~\citep{Bogner2003PLB_Vlowk, Bogner2003PR_VlowkReview} and the unitary correlation operator method~\citep{Feldmeier1998NPA_UCOM, Roth2010PPNP_UCOMReview} both implemented renormalization group transformations for nucleon-nucleon forces. 
Together with the development of chiral effective field theory,
these RG approaches revealed that nuclear forces did not need to be highly nonperturbative.
RG transformations could turn even very ``hard'' Hamiltonians like AV18
into phase-shift equivalent low-resolution interactions
that were more amenable to many-body calculations (as discussed above).
However, the RG transformations of nucleon-nucleon interactions also induce many-body forces, which must be accounted for,
and both methods faced difficulties with the consistent renormalization group transformations for three-nucleon forces.

This challenge was ultimately solved by the SRG, where consistently transforming nucleon-nucleon and three-nucleon potentials is relatively straightforward.
The framework to consistently transform nucleon-nucleon and three-nucleon potentials was first developed in a discrete harmonic oscillator basis in Ref.~\citep{Jurgenson2009PRL_SRG_ManyBody}
and after some time in a momentum-space basis as well~\citep{Hebeler2012PRC_SRG3N}.
Still, consistently SRG-transformed potentials have seen mixed success in nuclear structure calculations~\citep{Binder2018PRC_LENPICSMSNN, Hoppe2019PRC_ChiralMedMass, Huther2020PLB_EMNSRG, Soma2020PRC_SCGFlnl},
often facing substantial SRG resolution scale dependence and challenges in the simultaneous description of nuclei and nuclear matter.
This hints at the fact that induced four-nucleon forces may be significant
and thus nonnegligible for truly RG-invariant predictions.

\begin{figure}[t!]
    \centering
    \includegraphics[width=0.6\linewidth]{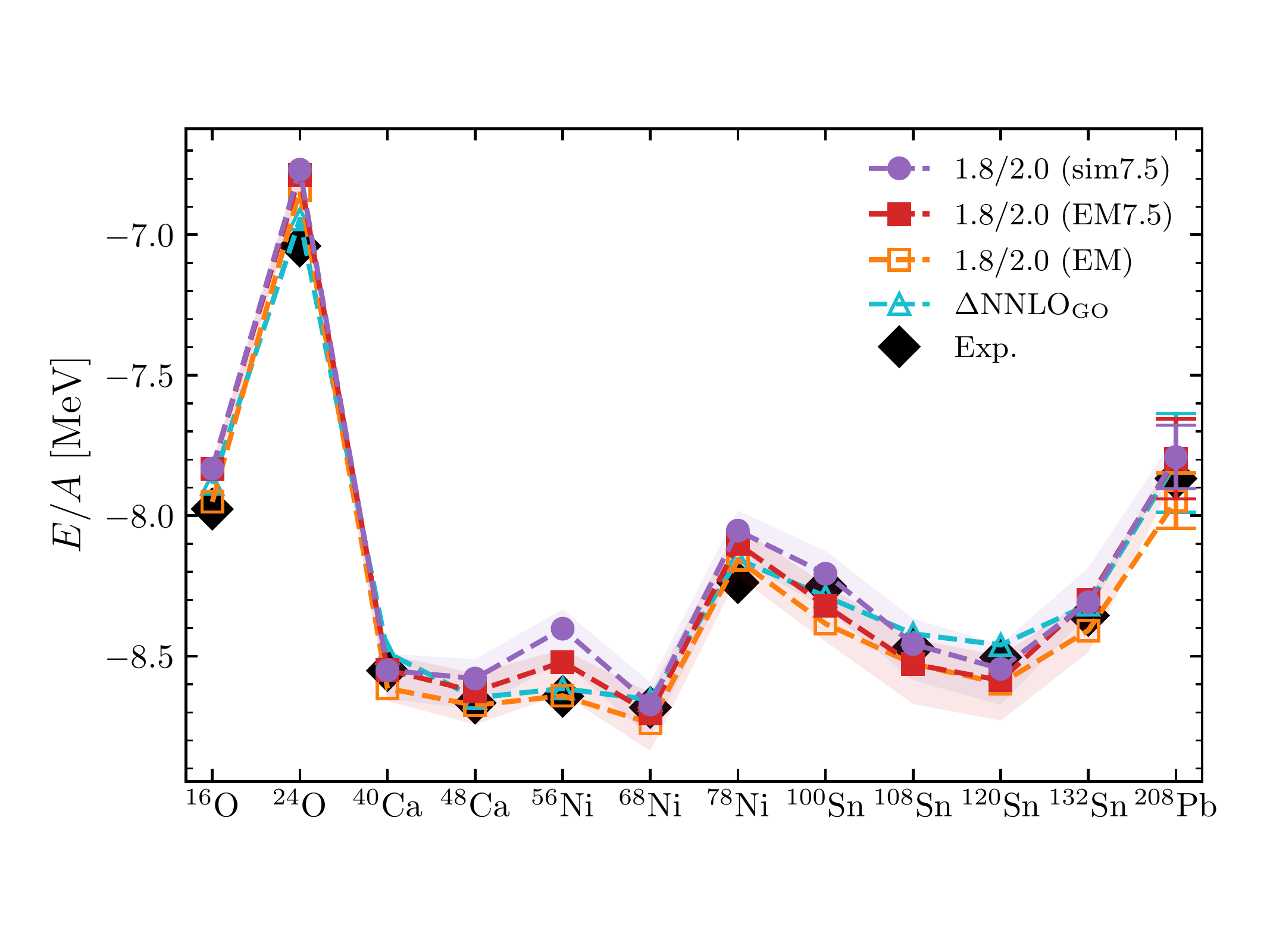}
    \caption{Ground-state energy per nucleon $E/A$ for selected nuclei computed using low-resolution nuclear Hamiltonians~\citep{Hebeler2011PRC_SRG3NFits, Jiang2020PRC_DN2LOGO, Arthuis2024arxiv_LowResForces}.
    The 1.8/2.0~(EM), 1.8/2.0~(EM7.5), and 1.8/2.0~(sim7.5) Hamiltonian are constructed using SRG-transformed potentials with $\lambda=1.8\:\fmi$.
    The $\Delta$NNLO$_\mathrm{GO}$ Hamiltonian is instead explicitly constructed with a low cutoff of $\Lambda=2.0\:\fmi$, motivated by the success of SRG-evolved low-resolution Hamiltonians.
    The agreement with experimental binding energies across the nuclear chart is remarkable.
    Figure adapted from~\cite{Arthuis2024arxiv_LowResForces}.}
    \label{fig:LowResolutionSummary}
\end{figure}

Still, there was a period where low-resolution nucleon-nucleon potentials proliferated
without consistently transformed three-nucleon interaction.
Around the same time, three-nucleon potentials from chiral effective field theory were developed~\citep{Epelbaum2002PRC_Chiral3NF}.
It was clear that both the initial three-nucleon forces along with the induced three-nucleon forces from the RG transformation would be required
(at least)
for RG-invariant calculations~\citep{Hebeler2010PRC_ChiralEFTNeutronMatter}.
A pragmatic approach emerged to address this problem without the tools required for a consistent SRG transformation:
Instead of transforming the nucleon-nucleon and three-nucleon potentials together,
one first used RG methods to transform nucleon-nucleon potentials to low resolution
and then refit of the resolution-dependent short-range three-nucleon force couplings 
to ensure RG invariance of three- and four-body observables~\citep{Hebeler2011PRC_SRG3NFits, Arthuis2024arxiv_LowResForces}.
The intuition behind this choice is that the RG transformation of nucleon-nucleon potentials leaves the long-range pion-exchange physics untouched
and primarily shifts repulsive short-range two-body interactions to short-range three-body interactions.
these induced short-range three-body terms would ideally be exactly captured by a consistent SRG transformation,
A refitting of the short-range three-nucleon force couplings is expected to approximately capture these induced short-range three-body terms
that would otherwise be exactly captured by a consistent SRG transformation.
This approach works to some degree,
but it has been shown that such Hamiltonians are less RG-invariant than consistently SRG-tranformed Hamiltonians~\citep{Hebeler2021PR_3NF}.
Despite this, some Hamiltonians constructed this way have been very successful in calculations of nuclei up to $^{208}$Pb
(see Fig.~\ref{fig:LowResolutionSummary}).

The success of low-resolution Hamiltonians constructed using RG methods has also motivated the use of low cutoffs in effective field theories.
Several established Hamiltonians employ low cutoffs of $\Lambda=394$--$450\:\mev$~\citep{Roth2012PRL_NO2B3N, Ekstrom2015PRC_N2LOsat, Soma2020PRC_SCGFlnl, Jiang2020PRC_DN2LOGO, Hu2022NP_Pb208},
which give them similar convergence properties to RG-transformed low-momentum Hamiltonians (see Fig.~\ref{fig:LowResolutionSummary}).
The tradeoff is that employing low cutoffs can degrade the reproduction of nucleon-nucleon scattering data at higher lab energies~\citep{Ekstrom2013PRL_N2LOopt, Carlsson2016PRX_NNLOsim, Jiang2020PRC_DN2LOGO, Epelbaum:2022cyo}.

Recently, in nuclear lattice effective field theory~\citep{Lee2009PPNP_LatticeReview},
wavefunction matching for nucleon-nucleon forces was developed~\citep{Elhatisari2024N_WaveFunctionMatching}
following established philosophies of RG transformations of nucleon-nucleon potentials.
Problematic short-range nucleon-nucleon interactions are transformed using a unitary transformation
and matched to long-range interactions from effective field theory.
The three-nucleon force parameters are then optimized to binding energies of light and medium-mass nuclei.
This approach applies the established philosophies of RG transformations of nucleon-nucleon potentials
(e.g., $V_{\mathrm{low}\:k}$, the unitary correlation operator method, and the SRG)
and of low-resolution fits of three-nucleon forces to lattice Hamiltonians.
The resulting Hamiltonians are amenable to the perturbative treatment used in lattice EFT calculations\footnote{
    Nuclear lattice effective field theory treats many subleading interactions in perturbation theory.
    Recent work has extended fully nonperturbative coupled cluster and in-medium similarity renormalization group calculations to lattice bases~\citep{Rothman2025PRC_NO2B, Rothman2025arxiv_NuLattice, Rothman:2026ixb}.
}
and have been able to provide a good description of bulk properties of nuclei up to $^{40}$Ca.

\section{Conclusion}

The similarity renormalization group and low-resolution nuclear forces continue to play key roles in nuclear theory today~\citep{Tews2022FS_CollectionOfPerspectives}.
Nuclear theory calculations are rapidly expanding their frontiers,
computing superheavy isotopes, nuclei at the drip lines, fundamental interactions in nuclei, reactions in exotic systems, and properties of neutron-star matter.
Improving our understanding of nuclear forces and developing more accurate low-resolution Hamiltonians
will be essential to continued progress.

\section*{Data availability}

Accompanying this chapter is a Python code \texttt{\detokenize{nn-srg}}~\citep{Heinz2025_NNSRGGit, Heinz2025_NNSRGZenodo} to allow readers to experiment with the SRG transformation of nucleon-nucleon potentials.
Users can explore the momentum-space SRG transformation of nucleon-nucleon potentials, featuring both the AV18 and EM500 potentials,
and test SRG scale invariance on deuteron binding energies and nucleon-nucleon phase shifts.
\texttt{\detokenize{nn-srg}} contains scripts to generate Figs.~\ref{fig:SRGContours} and~\ref{fig:SRGSuppression}.

\begin{ack}[Acknowledgments]

I am grateful to Scott Bogner, Dick Furnstahl, Kai Hebeler, Heiko Hergert, Robert Perry, Robert Roth, and Achim Schwenk
for instruction, discussions, and research projects on the similarity renormalization group
and to many more colleagues and collaborators for fruitful discussions on nuclear physics.
I thank Francesca Bonaiti, Dick Furnstahl, Ulrich Heinz, Thomas Papenbrock, and Daniel Phillips for valuable feedback on drafts of this chapter.

This work was supported
by the U.S.\ Department of Energy, Office of Science, Office of Advanced Scientific Computing Research and Office of Nuclear Physics, Scientific Discovery through Advanced Computing (SciDAC) program (SciDAC-5 NUCLEI)
and by the Laboratory Directed Research and Development Program of Oak Ridge National Laboratory, managed by UT-Battelle, LLC, for the U.S.\ Department of Energy.
This research used resources of the Oak Ridge Leadership Computing Facility located at Oak Ridge National Laboratory, which is supported by the Office of Science of the Department of Energy under contract No.~DE-AC05-00OR22725.

\end{ack}

\bibliographystyle{apsrev4-1-mh-mod}
\bibliography{noarxiv}

\end{document}